\documentclass[12pt,preprint]{aastex}
\usepackage{graphicx}

\title{A Jet Source of Event Horizon Telescope Correlated Flux in M87}
\author{Brian Punsly\altaffilmark{1}}
 \altaffiltext{1}{1415 Granvia Altamira, Palos
Verdes Estates CA, USA 90274 and ICRANet, Piazza della Repubblica 10
Pescara 65100, Italy, brian.punsly1@cox.net}
\begin{document}

\begin{abstract} Event Horizon Telescope (EHT) observations at 230 GHz are combined with Very Long Baseline Interferometry
(VLBI) observations at 86 GHz and high resolution Hubble Space
Telescope optical observations in order to constrain the broadband
spectrum of the emission from the base of the jet in M87. The recent
VLBI observations of Hada et al provide much stricter limits on the
86 GHz luminosity and component acceleration in the jet base than
was available to previous modelers. They reveal an almost hollow jet
on sub-mas scales. Thus, tubular models of the jet base emanating
from the innermost accretion disk are considered within the region
responsible for the EHT correlated flux. There is substantial
synchrotron self absorbed opacity at 86 GHz. A parametric analysis
indicates that the jet dimensions and power depend strongly on the
86 GHz flux density and the black hole spin, but weakly on other
parameters such as jet speed, 230 GHz flux density and optical flux.
The entire power budget of the M87 jet, $\lesssim
10^{44}\rm{ergs/sec}$, can be accommodated by the tubular jet. No
invisible, powerful spine is required. Even though this analysis
never employs the resolution of the EHT, the spectral shape implies
a dimension transverse to the jet direction of 12-21 $\mu \rm{as}$
($\sim$24-27$\, \mu \rm{as}$) for $0.99 > a/M
> 0.95$ ($a/M\sim 0.7 $), where $M$ is the mass and $a$ is the
angular momentum per unit mass of the central black hole.

\end{abstract}

\section{Introduction}
The Event Horizon Telescope (EHT) is a global Very Long Baseline
Interferometer (VLBI) that can achieve $\sim 25\, \mu\rm{as}$
resolution at 230 GHz \citep{kri15}. 86 GHz VLBI has far superior
imaging capabilities at the expense of lower resolution, $\sim 60\,
\mu\rm{as}$ \citep{kim16,aki17}. In this analysis, the capabilities
of EHT to describe the base of the jet in M87 is enhanced by
combining these observations with 86 GHz global VLBI and high
resolution optical observations with the Hubble Space Telescope
(HST). 86 GHz VLBI observations reveal a hollow jet on sub-mas
scales \citep{had16}. Thus, squat tubular models of the jet base
within the compact region producing the EHT correlated flux
(referred to as the EHT core, the EHTC) are studied in this article.
By constraining the broadband spectrum with 86 GHz VLBI (that
constrains the SSA opacity) and HST optical observations (that
constrain the high energy tail of the synchrotron power law) far
more information is revealed than considering the EHT observations
in isolation. This information allows for a determination of the
radius and the vertical magnetic flux of the jet base within the
models. Since the data is not of matched resolution, there is much
uncertainty in the exact broadband spectrum of the EHTC. One can
compensate for this by considering a wide range of plausible fits to
the observed data. This study considers 15 models that produce 9
fits to the data.
\par Section 2 introduces the tubular jet and the physics required to
describe the radiation from a jet located close to a rotating BH
(black hole). In Section 3, the broadband spectrum from the region
responsible for the EHTC is estimated. Then, various tubular jet
models are computed that are consistent with observation. This
facilitates a parameter study of the possible plasma state. It is
assumed that the mass of central BH is $M_{bh} = 6 \times 10^{9}
M_{\odot}$ or $M=8.86 \times 10^{14} \rm{cm}$ in geometrized units,
$\sim 3.5\mu\rm{as}$ at 16.7 Mpc \citep{geb11}.

\section{Tubular Jet Models}
This section describes physics relevant to the tubular jet model.
The basic model is a tubular geometry with an inner radius at the
ISCO (innermost stable orbit) and an outer radius, $R$, with a
height with a fiducial height $H=2R$ that is allowed to vary in some
of the models (see Figure 1). The rest frame evaluated number
density and vertical poloidal magnetic field that is anchored in the
equatorial plane ($N$ and $B^{P}$, respectively) are both constant
throughout the volume.

\subsection{Synchrotron Emission and Absorption} The underlying power law for the flux density is defined
by $F_{\nu}(\nu= \nu_{o}) = F\nu_{o}^{-\alpha}$, where $\alpha$ is
the spectral index and $F$ is a constant. Observed quantities will
be designated with a subscript, ``o", in the following expressions.
The SSA attenuation coefficient in the plasma rest frame, noting
that the emitted frequency is designated by $\nu$, is given
in{Equation (3.33) of \citet{gin69},
\begin{eqnarray}
&& \mu(\nu)=\overline{g(n)}\frac{e^{3}}{2\pi
m_{e}}N_{\Gamma}(m_{e}c^{2})^{2\alpha} \left(\frac{3e}{2\pi
m_{e}^{3} c^{5}}\right)^{\frac{1+2\alpha}{2}}\left(B\right)^{(1.5
+\alpha)}\left(\nu\right)^{-(2.5 + \alpha)}\;,\\
&& \overline{g(n)}= \frac{\sqrt{3\pi}}{8}\frac{\overline{\Gamma}[(3n
+ 22)/12]\overline{\Gamma}[(3n + 2)/12]\overline{\Gamma}[(n +
6)/4]}{\overline{\Gamma}[(n + 8)/4]}\;, \\
&& N=\int_{\Gamma_{min}}^{\Gamma_{max}}{N_{\Gamma}\Gamma^{-n}\,
d\Gamma}\;,\; n= 2\alpha +1 \;,
\end{eqnarray}
where $\Gamma$ is the ratio of lepton energy to rest mass energy,
$m_{e}c^2$ which should be distinguished from $\overline{\Gamma}$
which is the gamma function. $B$ is the magnitude of the total
magnetic field. The power law spectral index for the flux density is
$\alpha=(n-1)/2$. The low frequency VLBI data do not strongly
constrain the low energy cutoff, $E_{min} = \Gamma_{min}m_{e}c^2$,
due to insufficient spatial resolution. Thus, no results that depend
on $\Gamma_{min}$ are discussed in this paper. There are still many
interesting conclusion that can be drawn from the models. The high
energy cutoff, $E_{max} = \Gamma_{max}m_{e}c^2$ might be revealed by
the UV HST observations (see Section 3.1), but this information is
never needed in the analysis. The conversion to the observer's
frequency, $\nu_{o}$, is given by $\nu = \nu_{o} / \delta$, where
$\delta$ is the total Doppler factor that includes gravitational
redshift and relative motion. The SSA opacity in the observer's
frame, $\mu(\nu_{o})$, is obtained by direct substitution of $\nu =
\nu_{o} / \delta$ into Equation (1). The homogeneous approximation
yields a simplified solution to the radiative transfer equation for
the intensity, $I_{\nu}$, from the SSA source \citep{gin65}
\begin{eqnarray}
&& I_{\nu}(\nu) = \frac{j_{\nu}(\nu)}{\mu(\nu)} \times \left(1 -
e^{-\mu(\nu)L}\right)\;,
\end{eqnarray}
where $L$ is the distance traversed by the radiation through the
plasmoid and the synchrotron emissivity is given in \citet{tuc75} as
\begin{eqnarray}
&& j_{\nu} = 1.7 \times 10^{-21} [4 \pi N_{\Gamma}]a(n)B^{(1
+\alpha)}\left(\frac{4
\times 10^{6}}{\nu}\right)^{\alpha}\;,\\
&& a(n)=\frac{\left(2^{\frac{n-1}{2}}\sqrt{3}\right)
\overline{\Gamma}\left(\frac{3n-1}{12}\right)\overline{\Gamma}\left(\frac{3n+19}{12}\right)
\overline{\Gamma}\left(\frac{n+5}{4}\right)}
       {8\sqrt\pi(n+1)\overline{\Gamma}\left(\frac{n+7}{4}\right)} \;.
\end{eqnarray}
One can transform this to the observed flux density, $S(\nu_{o})$,
in the optically thin region of the spectrum (for M87 in the IR and
optical) using the relativistic transformation relations from
\citet{lin85},
\begin{eqnarray}
 && S(\nu_{o}) = \frac{\delta^{(3 + \alpha)}}{4\pi D_{L}^{2}}\int{j_{\nu}^{'} d V{'}}\;,
\end{eqnarray}
where $D_{L}$ is the luminosity distance and in this expression, the
primed frame is the rest frame of the plasma.

\begin{figure}
\begin{center}
\includegraphics[width= 0.85\textwidth]{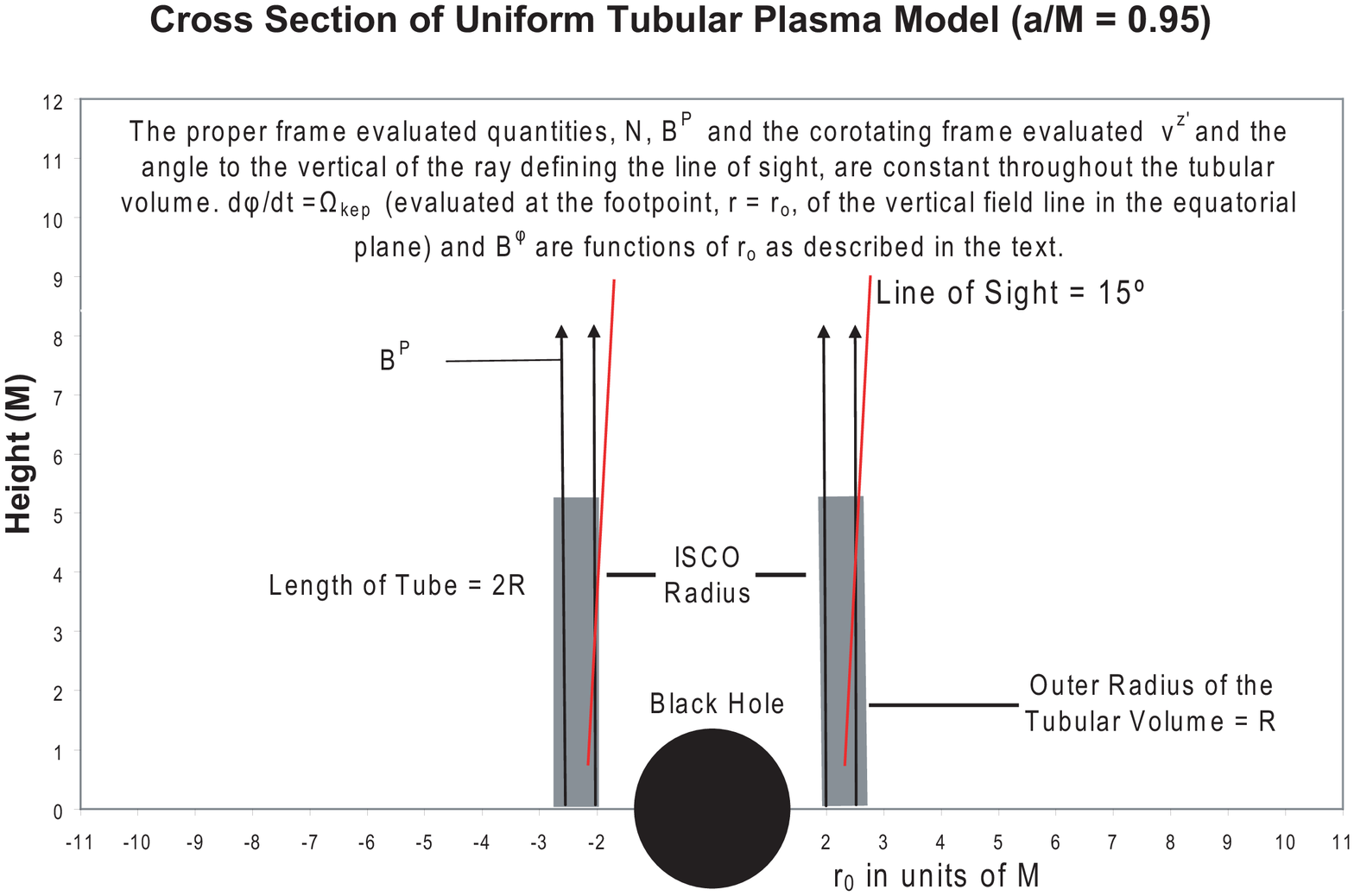}
\caption{The details of the uniform tubular plasmoid model. The
example chosen here is the fiducial model with $H/R=2$ and
$a/M=0.95$.}
\end{center}
\end{figure}
\begin{table}
\caption{Models of the Jet Base}
{\footnotesize\begin{tabular}{cccccccc} \tableline\rule{0mm}{3mm}
Model &  230 GHz & 86 GHz & $3.72 \times 10^{14}$ Hz & $\alpha$ &$a/M$ & $v^{z'}/c$ & Jet Length\\
         &  $F_{\nu}$(mJy) &  $F_{\nu}$(mJy) &  $F_{\nu}$(mJy) & &            &        & ($H$)\\
\tableline \rule{0mm}{3mm}
A & 830 & 450 & 0.5 &1.01& 0.99  & 0.1 & $2R$ \\
B & 830 & 450 & 0.5 &1.01& 0.95 & 0.1 & $2R$ \\
C & 830 & 450 & 0.5 &1.01& 0.70 & 0.1 & $2R$  \\
D & 830 & 450 & 0.5 &1.01& 0.95 & 0.05 & $2R$   \\
E & 830 & 450 & 0.5 &1.01& 0.95 & 0.2 & $2R$  \\
F & 830 & 450 & 0.75 &0.96& 0.95 & 0.1 & $2R$ \\
G & 830 & 450 & 0.37 &1.06& 0.95 & 0.1 &$2R$ \\
H & 830 & 550 & 0.5 &1.01& 0.95 & 0.1 & $2R$ \\
I & 830 & 350 & 0.5 &1.02& 0.95 & 0.1 & $2R$ \\
J & 980 & 450 & 0.5 &1.04& 0.95 & 0.1 & $2R$ \\
K & 630 & 450 & 0.5 &0.98& 0.95 & 0.1 & $2R$ \\
L & 500 & 350 & 0.5 &0.94& 0.95 & 0.1 & $2R$\\
M & 330 & 350 & 0.5 &0.88& 0.95 & 0.1 & $2R$\\
N & 820 & 540 & 0.5 &1.01& 0.95  & 0.1 & $4R$ \\
O & 820 & 560 & 0.5 &1.01& 0.95 & 0.1 & $8R$ \\
\end{tabular}}

\end{table}

\subsection{Relativistic Considerations} Calculations are computed on
the background of the Kerr metric (that of a rotating uncharged BH),
with mass, $M$, and angular momentum per unit mass, $a$. In
Boyer-Lindquist coordinates, $g_{\mu\nu}$, is given by the line
element
\begin{eqnarray}
 && d s^{2} \equiv g_{\mu\nu}\, dx^{\mu} dx^{\nu}=-\left (1-\frac{2Mr}{\rho^{2}}\right) d t^{2} +\rho^{2} d\theta^{2}\nonumber\\
&& +\left (\frac{\rho^{2}}{\Delta}\right) dr^{2} -\frac{4Mra}{\rho^{2}}\sin^{2}\theta\nonumber\\
&&  d\phi \,dt+\left[(r^{2}+a^{2})+\frac{2Mra^{2}}{\rho^{2}}\sin^{2}
\theta\right ] \sin^{2} \theta \, d\phi^{2} \; ,
\end{eqnarray}
where, $\rho^{2}=r^{2}+a^{2}\cos^{2}\theta$ and $\Delta =
r^{2}-2Mr+a^{2}$. The event horizon is defined by
$r_{_{+}}=M+\sqrt{M^{2}-a^{2}}$.  The magnetic field, $B^{P}$, is vertical and
anchored in the equatorial plane. It is assumed to rotate with an
angular velocity as viewed from asymptotic infinity, $\Omega_{F}
\approx \Omega_{\rm{kep}}$, the Keplerian angular velocity at the
foot-point (at $r=r_{o}$) of the field lines in the equatorial
plane:
\begin{equation}
\Omega_{\rm{kep}}(r_{o}) = \frac{M^{0.5}}{r_{o}^{1.5}+aM^{0.5}}\;.
\end{equation}
It should be noted in the following that the system might rotate
slightly slower due to magnetic torques. Each field line is defined
as vertical in terms of the Boyer-Lindquist coordinate system. There
is a constant coordinate displacement from the vertical axis that is
expressed by $r\sin{\theta}=r_{o}$.

It is useful to define an orthonormal ``co-rotating" frame
(co-rotates with the foot point, designated with a prime) for ease
of calculation with a 4-velocity
\begin{eqnarray}
&& e_0^{'} = \alpha_{\rm{kep}} ^{-1}\left(\frac{\partial}{\partial t} + \Omega_{\rm{kep}}(r_{o}) \frac{\partial}{\partial\phi}\right)\;, \nonumber \\
&& \alpha_{\rm{kep}}= \sqrt{-g_{t\, t} -2
\Omega_{\rm{kep}}(r_{o})g_{\phi\, t} - \Omega_{\rm{kep}}(r_{o})^{2}
g_{\phi\, \phi}}\; ,
\end{eqnarray}
where $\alpha_{\rm{kep}}$ is the gravitational redshift of the
co-rotating frame with respect to the stationary frames at
asymptotic infinity, For global calculations, we use the
hypersurface orthogonal, orthonormal ZAMO frames
\begin{eqnarray}
&& \hat{e}_0 = \alpha_{\rm{Z}} ^{-1}\left(\frac{\partial}{\partial t} + \Omega_{\rm{Z}}\frac{\partial}{\partial\phi}\right) \;,\; \Omega_{\rm{Z}} =\frac{-g_{\phi\, t}}{g_{\phi\phi}} \;,\; \alpha_{\rm{Z}} =\frac{\sqrt{\Delta} \sin{\theta}}{\sqrt{g_{\phi\phi}}}\;, \nonumber \\
&&\hat{e}_\phi=\frac{1}{\sqrt{g_{\phi\phi}}}\frac{\partial}{\partial\phi}\
,\;\hat{e}_r = \left( \frac{\Delta^{1/2}}{\rho} \right)
\frac{\partial}{\partial r} \ , \;\hat{e}_\theta =\left(
\frac{1}{\rho} \right) \frac{\partial}{\partial \theta} \; .
\end{eqnarray}
The boost to the orthonormal corotating frame is
\begin{equation}
v_{\rm{Kep_Z}}^{\phi}
=[\Omega_{\rm{kep}}(r_{o})-\Omega_{\rm{Z}}]\sqrt{g_{\phi\phi}}/\alpha_{\rm{Z}}\;,\;
\gamma_{\rm{Kep_Z}} =\alpha_{\rm{Z}}/\alpha_{\rm{kep}}
=\left[1-(v_{\rm{Kep_Z}}^{\phi}/c)^{2}\right]^{-0.5}\;,
\end{equation}
i.e., $\gamma_{\rm{Kep_Z}}B^{P'}=B^{P}_{\rm{ZAMO}}$ and in Equation
(7), $dV^{'}\approx \gamma_{\rm{Kep_Z}} dV_{\rm{ZAMO}}$.
\par Consider the frozen-in condition applied to the toroidal
magnetic field in the ZAMO frame,
\begin{equation} B^{\phi}_{\rm{ZAMO}}/B^{P}_{\rm{ZAMO}} = (v^{\phi}_{\rm{ZAMO}}-v^{\phi}_{F})/v^{z}_{\rm{ZAMO}}\;,
\end{equation}
 where $v^{\phi}_{F}$ is the
azimuthal velocity of the field, and $v^{\phi}_{\rm{ZAMO}}$ and
$v^{z}_{\rm{ZAMO}}$ are the azimuthal velocity and vertical velocity
of the bulk flow of plasma \citep{pun08}. In the region of jet
initiation, $v^{z}_{\rm{ZAMO}}$ is considered to be
non-relativistic. This is motivated theoretically as a boundary
condition in the equatorial plane and VLBI observations that
indicate apparent velocities, $v_{\rm{app}}=0.1 -0.4$c, $<2$ mas
from the core. Furthermore, $v_{app}$ increases from $\sim 0.15$c to
$ 1.5 - 2.0$c in the first 10 mas indicating strong magnetic forces,
requiring the jet base to be magnetically dominated
\citep{mer16,had16,had17}. In this Poynting flux dominated regime,
the conservation of angular momentum condition provides a constraint
on the azimuthal magnetic field,
\begin{equation}
\alpha_{\rm{Z}}\sqrt{g_{\phi\phi}} B^{\phi}_{\rm{ZAMO}} \approx
-\frac{\Omega_{F} \Phi}{k_{F}c} = \rm{constant}\;,
\end{equation}
where $\Phi$ is the total poloidal flux contained within the
cylindrical radius and $k_{F}$ is a geometrical factor that equals
$\pi$ for the assumed unform cylindrical asymptotic jet
\citep{pun08}. From Equations (8), (11) and (14),
\begin{equation}
B^{\phi}_{\rm{ZAMO}} \approx-\frac{\Omega_{\rm{kep}}(r_{o})
\Phi}{\Delta^{1/2}\sin{\theta} \pi c}\;.
\end{equation}
From Equations (13) - (15) and assuming that $v^{z} \sim 0.1c$, the
angular velocity of the plasma as viewed from asymptotic infinity,
$\Omega_{p} =d\phi/dt$, in the tubular plasmoid is in approximation
co-rotation with the field lines,
\begin{equation}
\mid \frac{\Omega_{p}- \Omega_{\rm{kep}}(r_{o})}
{\Omega_{\rm{kep}}(r_{o})}\mid\approx \mid-(v^{z}_{\rm{ZAMO}}/c)
\alpha_{\rm{Z}} \frac{SA_{\perp}}{\pi (g_{\phi\, \phi}\Delta)^{1/2}}
\mid \sim \mid-0.1 \alpha_{\rm{Z}}\mid \ll 1\;.\,
\end{equation}
where $SA_{\perp}$ is the cross sectional area of the plasmoid.
\begin{figure}

\includegraphics[width= 0.55\textwidth]{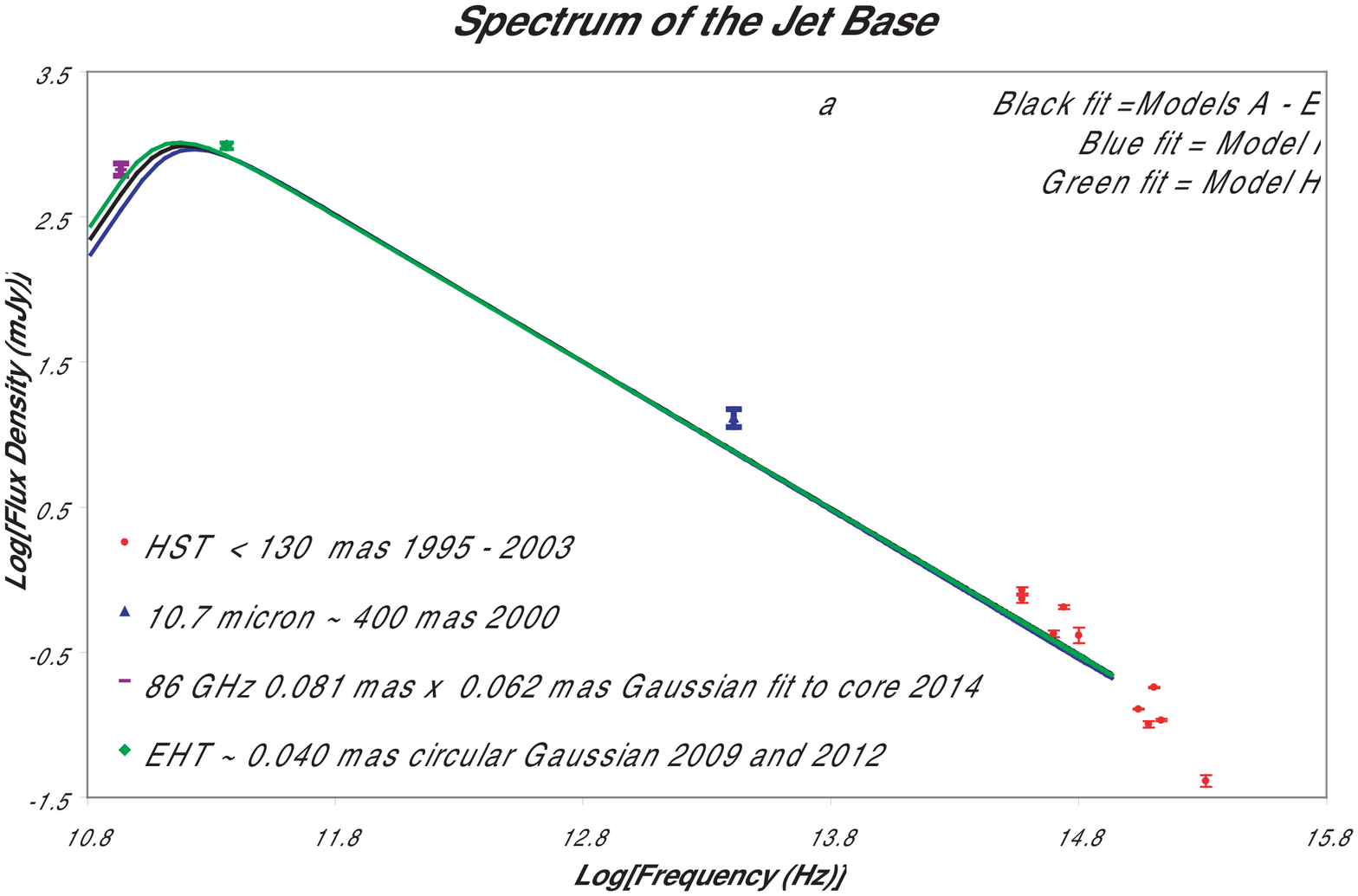}
\includegraphics[width= 0.55\textwidth]{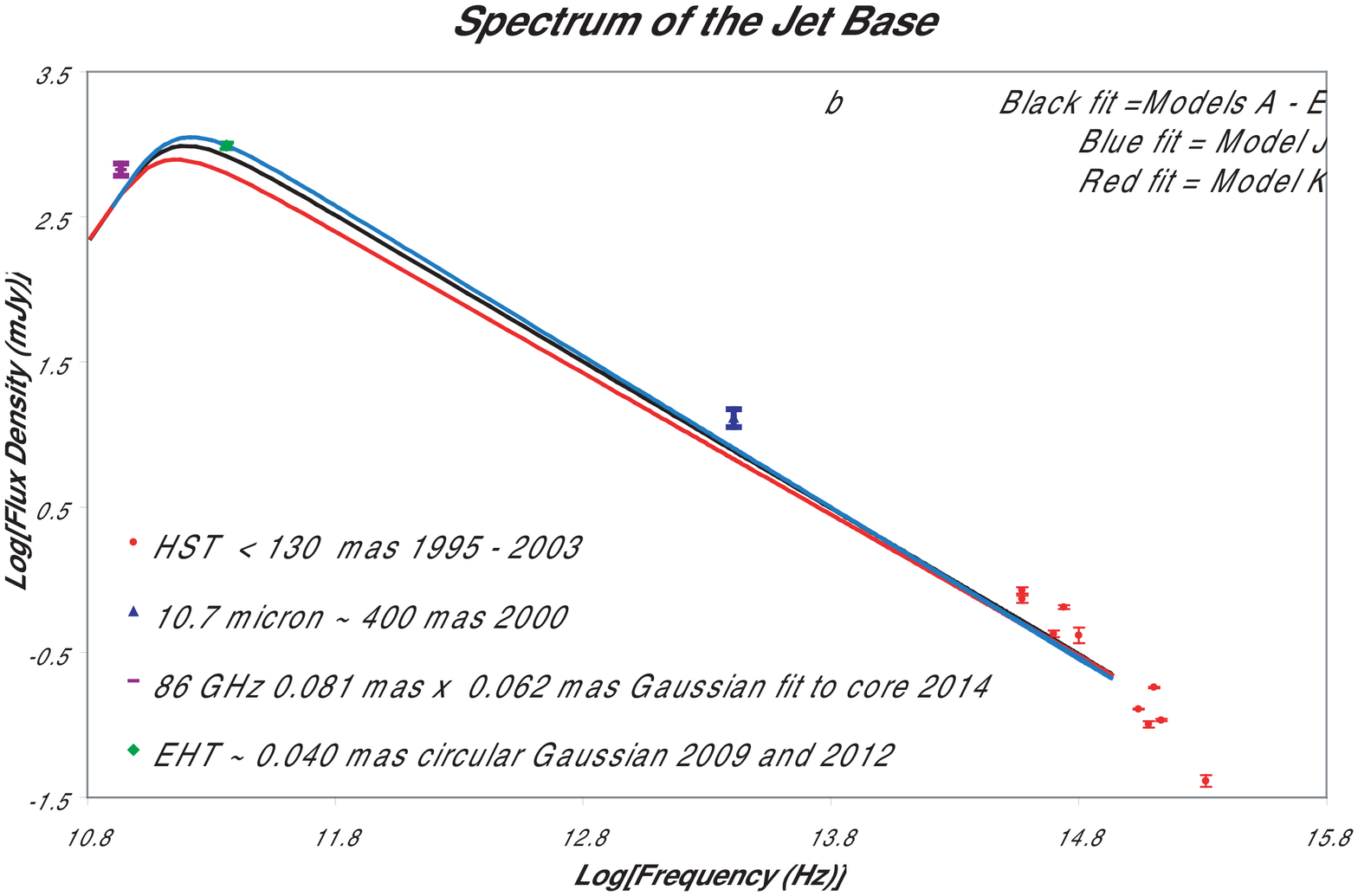}
\includegraphics[width= 0.55\textwidth]{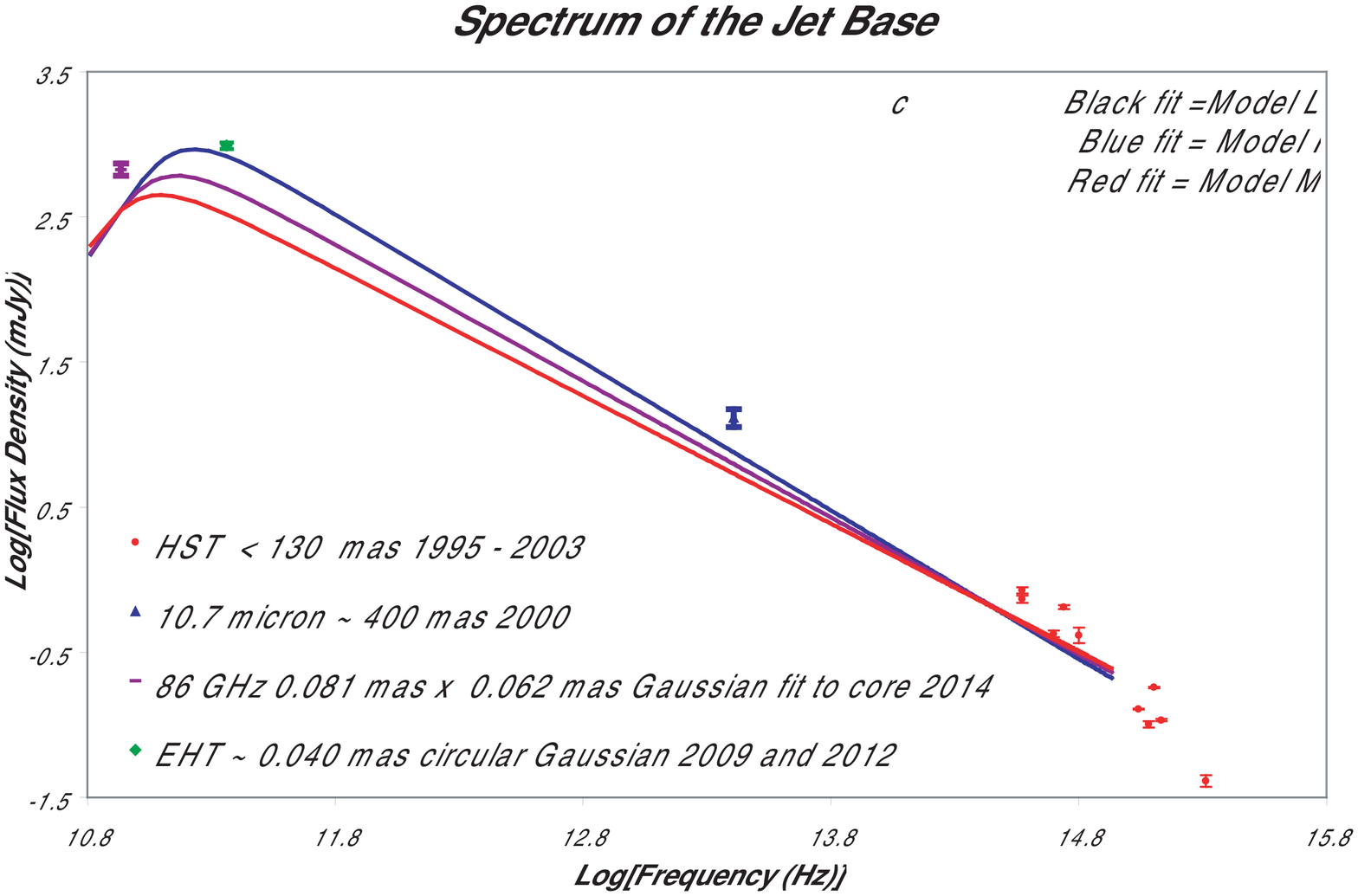}
\includegraphics[width= 0.55\textwidth]{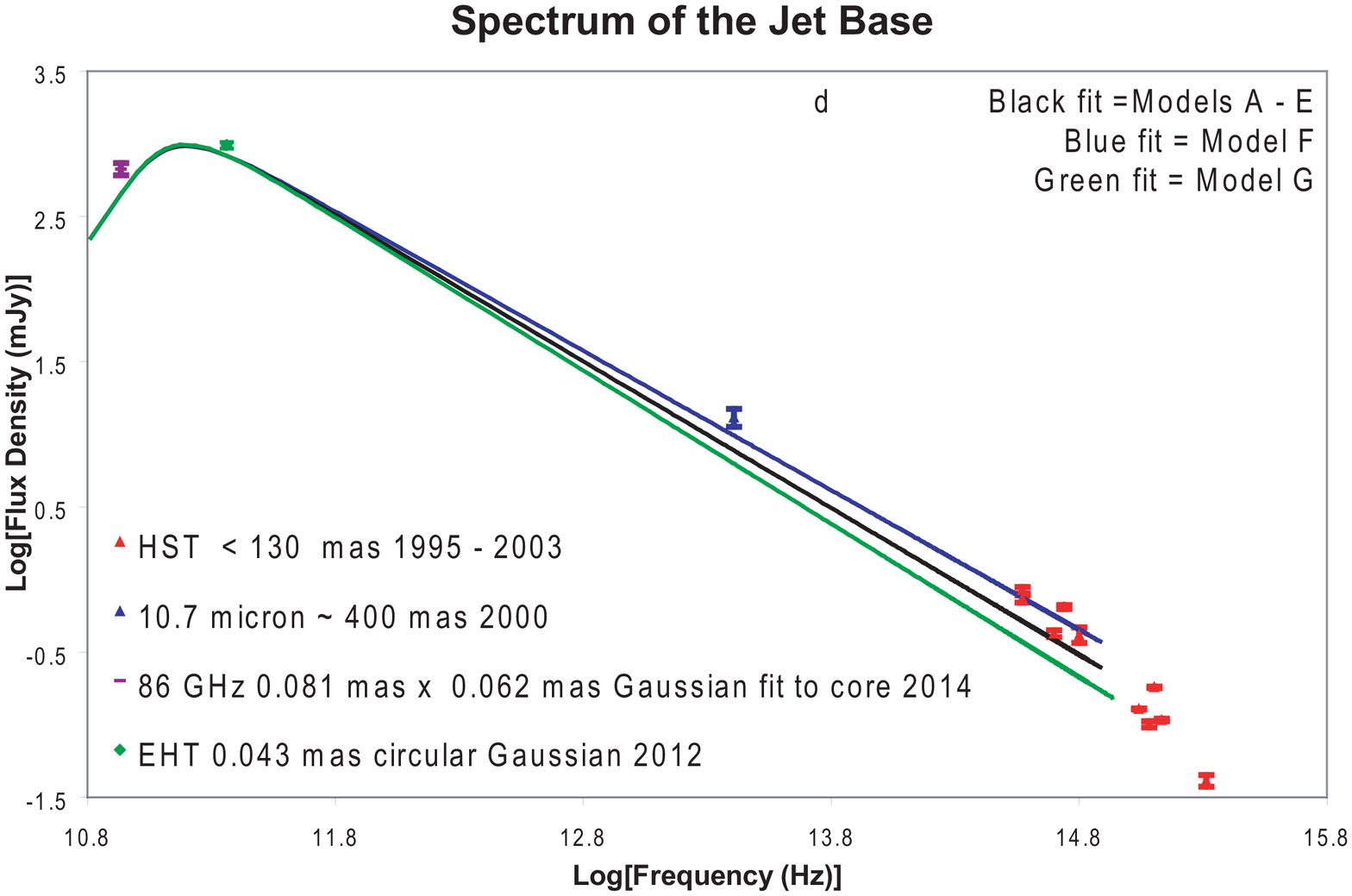}

\caption{The 13 distinct family of models in Table 1 (Models A - M),
produce 9 distinct spectral fits to the data. The fits are displayed
in 4 frames for clarity. Frame a (top left) compares different
choices for $F_\nu(\nu_{o} = 86 \rm{GHz})$ with $F_\nu( \nu_{o} =
230 \rm{GHz})$ held constant.. Frames b and c compare different
choices for $F_\nu(\nu_{o} = 230 \rm{GHz})$ with $F_\nu(\nu_{o} = 86
\rm{GHz})$ held constant. Frame d explores variations in the
strength of the synchrotron tail (note the spectral break in the
UV).}
\end{figure}
\par The Doppler factor is computed in a two step process \citep{lig75}. Consider a plasmoid that is moving along the vertical axis
(perpendicular to the plane of rotation) with a velocity $v^{z'}$ as
measured in the corotating frame and a bulk Lorentz factor,
$\gamma^{'}$. First, compute the Doppler shift in the co-rotating
frame due to relative motion if a photon is emitted from the
plasmoid at an angle, $\psi$, along the line of sight (LOS), where
$\psi$ is measured relative to the direction of bulk motion in the
co-rotating frame. Second, consider the gravitational redshift of
co-rotating frame in Equation (10) to find
\begin{equation}
\delta = \alpha_{\rm{kep}}/[\gamma^{'} (1-(v^{z'}/c)\cos{\psi})]\;.
\end{equation}
In the following, we will assume a value of $\psi = 15^{\circ}$
which is consistent with most of the observed jet motion
\citep{sta06}. This exact value does not affect the results
significantly because the putative plasma flow is subrelativistic.
\par The MHD Poynting flux in the magnetically dominated limit is
\begin{equation}
\int{S^{P}\mathrm{d}A_{_{\perp}}} =
k\frac{\Omega_{F}^{2}\Phi^{2}}{2\pi^{2} c},
\end{equation}
where $k$ is a geometrical factor that equals 1 for a uniform highly
collimated jet \citep{pun08}.

\begin{figure}
\includegraphics[width= 0.85\textwidth]{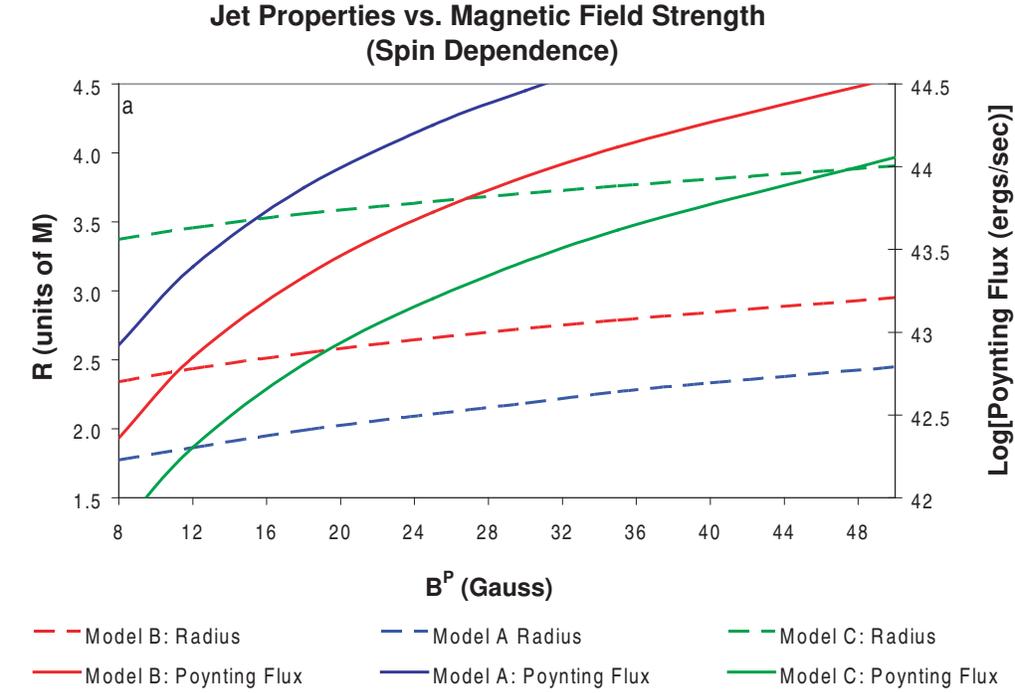}
\includegraphics[width= 0.85\textwidth]{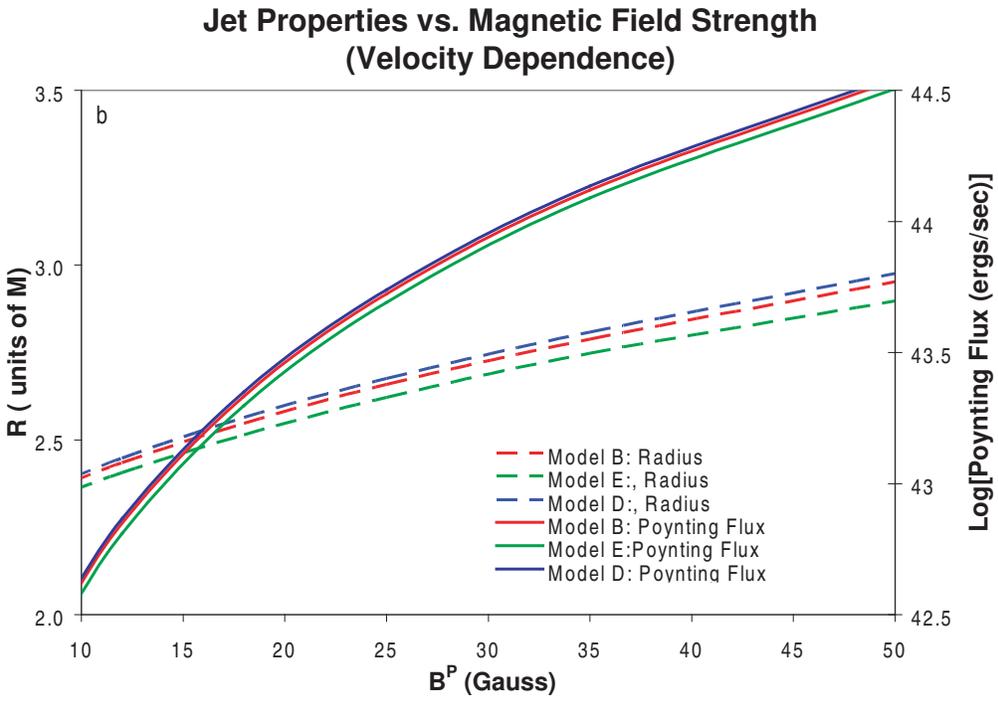}
\caption{An exploration of the different physical parameters that
produce the fiducial fit in Table 1 and Figure 2 as a consequence of
varying the assumptions on  $a$ and $v^{z'}$. The top frame shows
the dependence on spin. The bottom frame shows the minor effect of
varying the velocity, $v^{z'}$.}
\end{figure}

\par The simple parametric form of the spectrum in Equation (4) cannot be
produced by a homogeneous distribution of plasma due to the spatial
gradients in the gravitational redhift in the expression for
$\delta$ in Equation (17). The value of $\alpha_{\rm{kep}}$ varies
throughout the volume. Homogeneity is regained by implementing a
single value of $\alpha_{\rm{kep}}^{(2.5 + \alpha)}$ that occurs in
Equation (1) when it is written in terms of $\nu_{o}$ and a single
value of $\gamma_{\rm{Kep_Z}}\alpha_{\rm{kep}}^{(3 + \alpha)}$ in
Equation (7) throughout the compact calculational volume. Employing
the volumetric average of these values instead of the exact
coordinate dependent values is a major simplifying approximation
used in the calculation. The volumetric average of $B^{\phi}$ is
also implemented in the computation of $\mu$ and $j_{\nu}$. These
averages and constant plasma parameters result in simple radiative
transfer solutions as in Equation (4).
\section{Constructing Models of the Jet Base} In this section, the
formalism descried in the last section is used to construct tubular
models of the jet base. The first subsection describes the
observational data that is used to constrain the models. The second
subsection describes the models in detail and the resultant plasma
state of the jet base. Table 1 describes the parametric analysis of
various fits to the data and BH states.

\subsection{Constraining the Broadband Spectrum} In this section, the broadband spectrum of the jet base is
constrained. There are three portions of the spectrum to consider,
the EHT data that is located near the peak of the spectrum, the 86
GHz VLBI data that constrains the SSA opacity and the high frequency
synchrotron tail is constrained by HST high resolution optical/IR
photometry. The 86 GHz and optical data are of much lower resolution
and are considered as upper limits on the flux from the jet base.

\begin{figure}
\includegraphics[width= 0.55\textwidth]{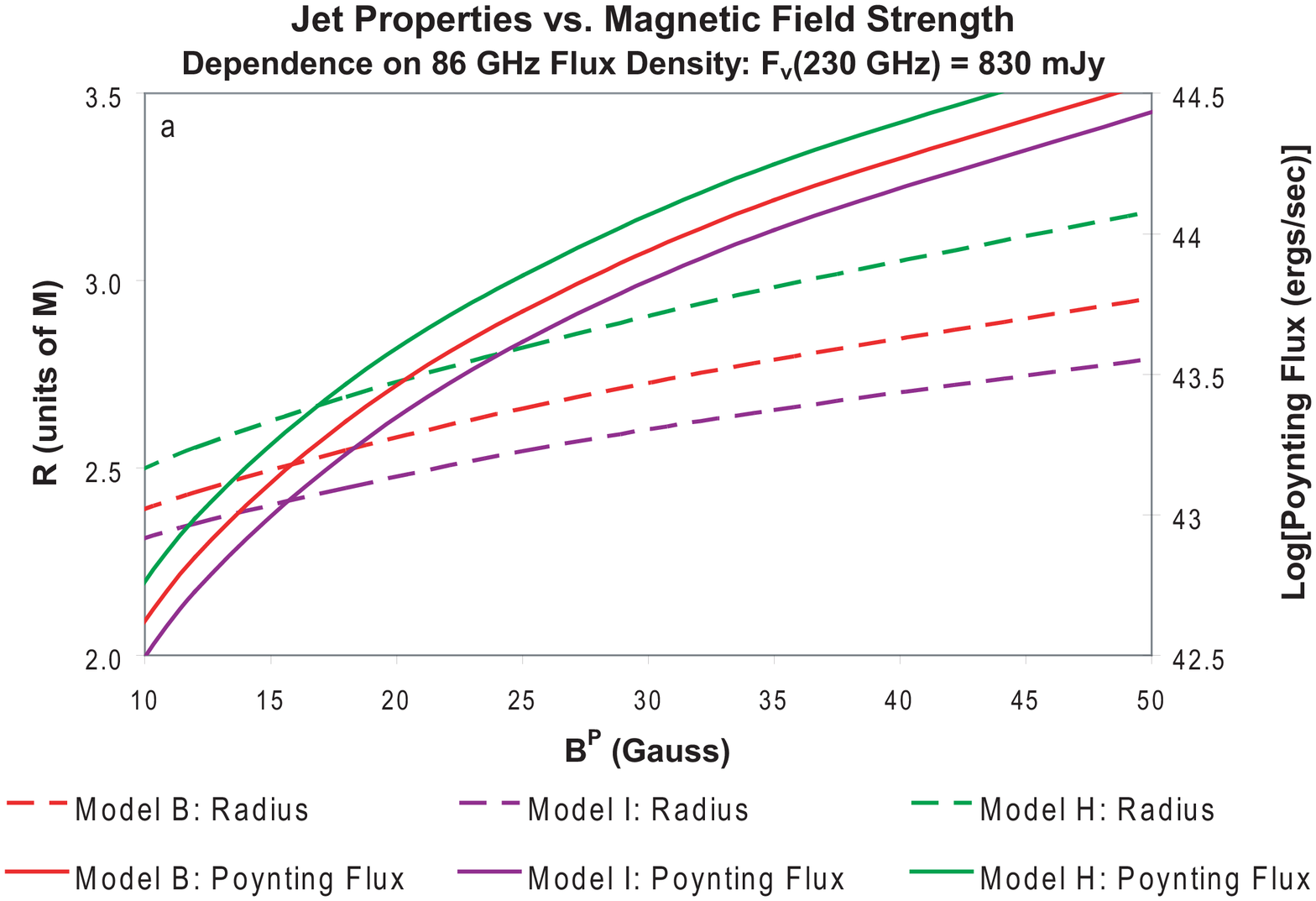}
\includegraphics[width= 0.55\textwidth]{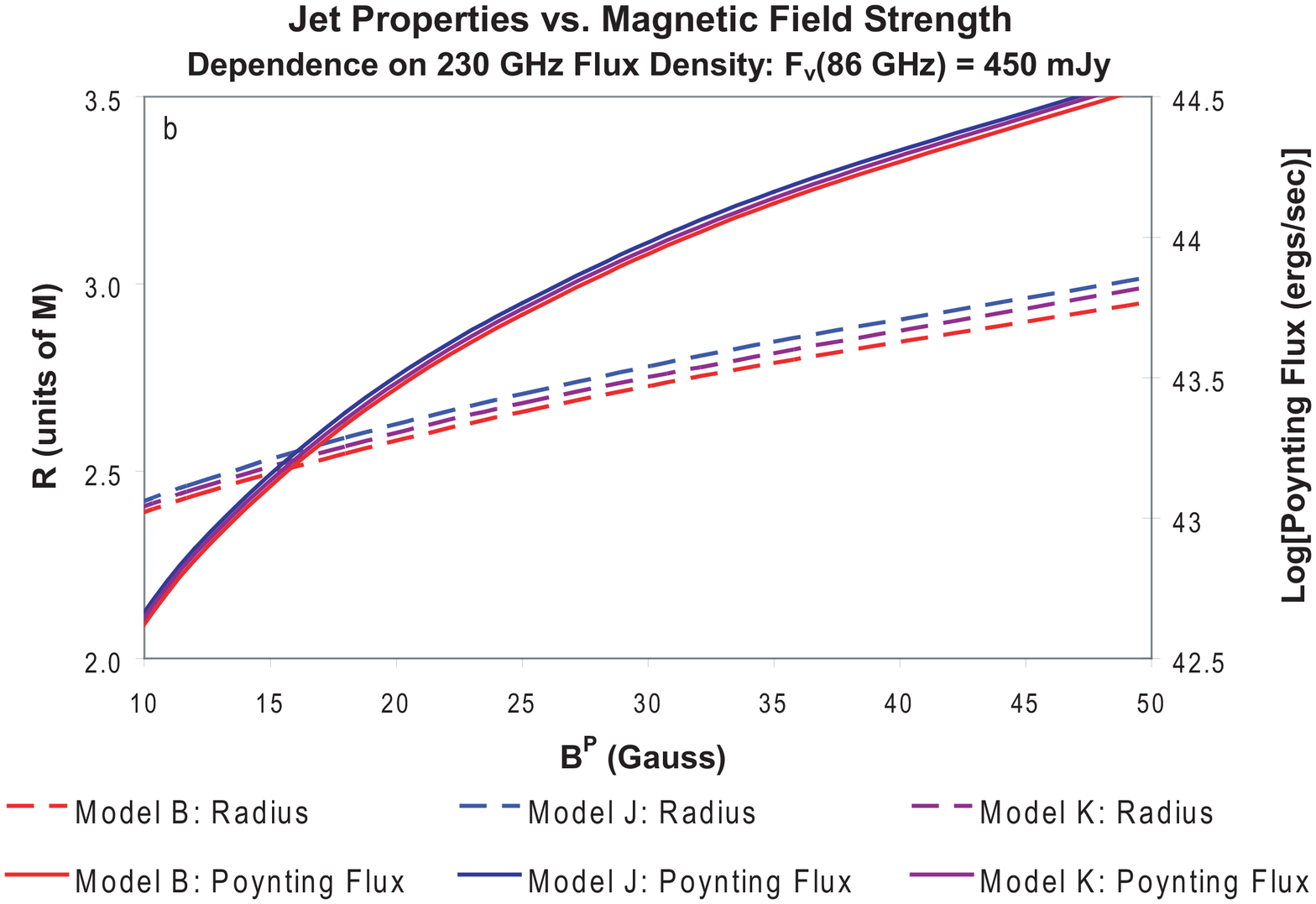}
\includegraphics[width= 0.55\textwidth]{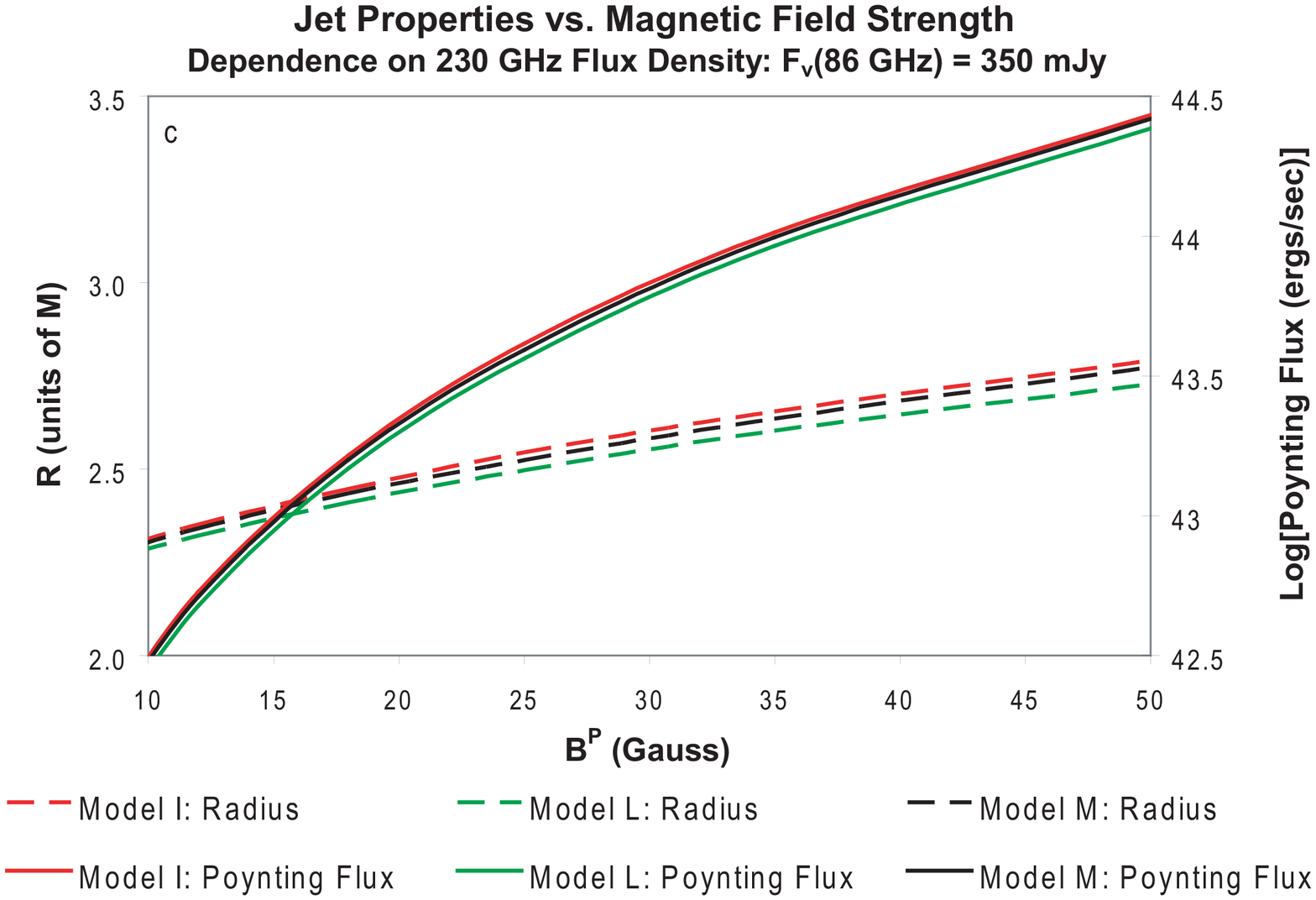}
\includegraphics[width= 0.55\textwidth]{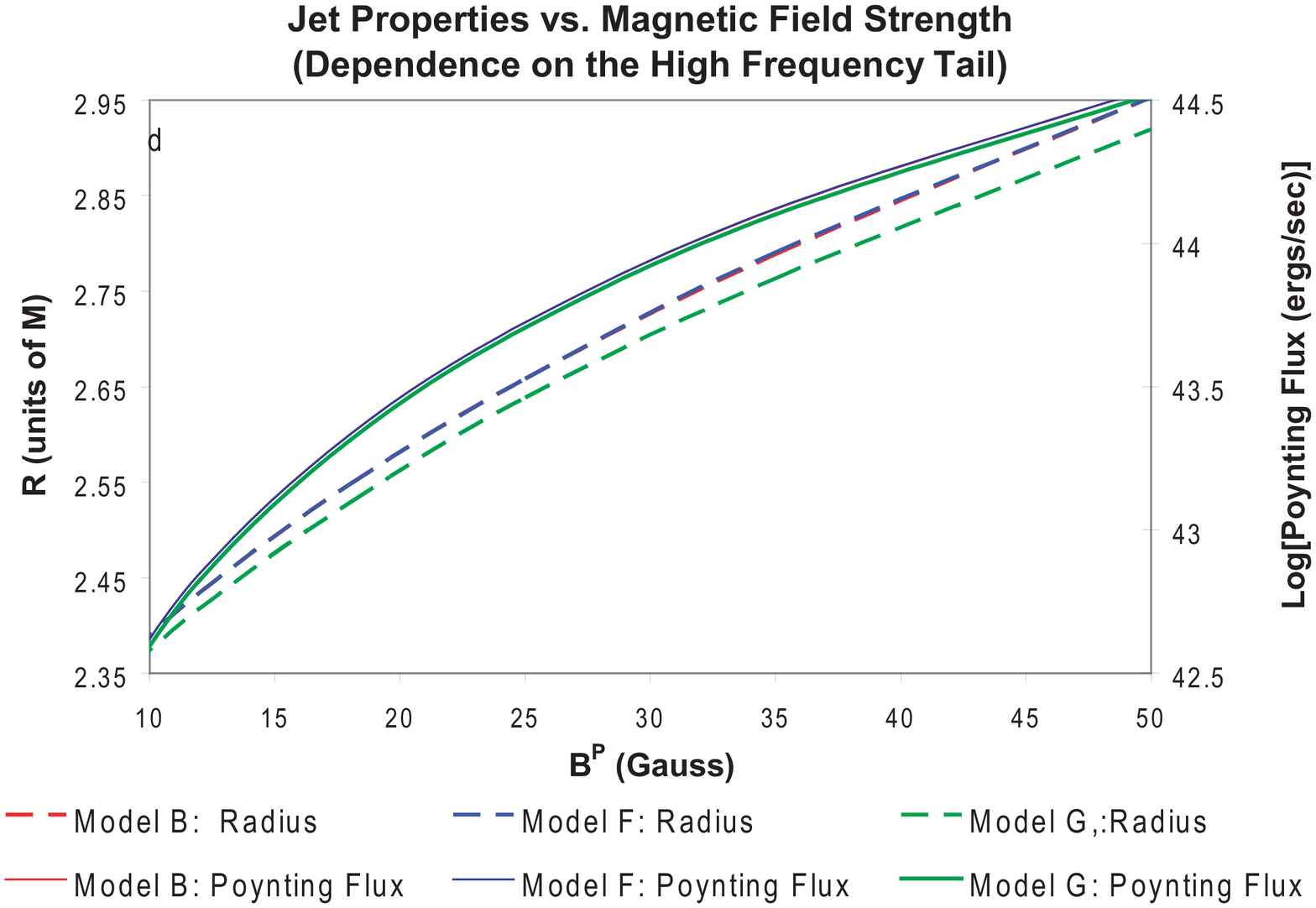}

\caption{Variations in the physical model are explored with $a$ and
$v^{z'}$ held fixed, but the fit is varied per Table 1 and Figure 2.
Frame a illustrates the effect of varying $F_\nu(\nu_{o} = 86
\rm{GHz})$ with $F_\nu(\nu_{o} = 230 \rm{GHz})$ held constant.
Frames b and c show the effects of varying $F_\nu(230 \rm{GHz})$ and
frame d the effects of varying the optical flux (a,b,c and d
correspond to the labels of the fits in Figure 2 as well). The only
significant variations occur when $F_\nu(\nu_{o} = 86\rm{GHz})$ is
changed.}

\end{figure}

There have been two published EHT detections of correlated flux of
M87 that are shown in Figure 2. The observations in 2009 and 2012
were fit with $980 \pm 40$ mJy and $980 \pm 50$ mJy of correlated
flux within a circular Gaussian component of FWHM (Full Width at
Half Maximum) of $40 \mu \rm{as}$ and $43 \mu \rm{as}$, respectively
\citep{doe12,aki15}. An exact value of correlated flux is not
utilized in the following (see Section 3.2).

\par The only published fit to the core with 86 GHz VLBI is 669 mJy in an
elliptical Gaussian fit 0.081 mas $\times$ 0.062 mas from 2014
observations \citep{had16}. These dimensions are larger than the EHT
fit and the observation is not contemporaneous. Furthermore, it is
not clear if it arises from the same region as the EHTC due to SSA
opacity. There is likely a significant fraction of $F_{\nu}(\nu_{o}
= 86 \, \rm{GHz})$ located within the EHTC since an extrapolation of
the SSA core shift analysis of \citet{had11} indicates that the EHTC
is only $\sim 10 \mu \rm{as}$ from the 86 GHz core. The precise
fraction of $F_{\nu}(\nu_{o} = 86 \, \rm{GHz})$ within the EHTC
cannot be determined as long as the 86 GHz VLBI baselines are
restricted to Earth. Thus, various values are chosen in the 9 fits
in Figure 2 and Table 1 in order to explore the dependence on the
plasma composition at the base of the jet on $F_{\nu}(\nu_{o} = 86
\, \rm{GHz})$. Due to the uncertainty, 669 mJy is considered only as
a crude upper limit from which to start the parametric variation of
$F_{\nu}(\nu_{o} = 86 \, \rm{GHz})$ from the EHTC.

\par The highest resolution HST optical/UV imaging achieves a resolution of
$\sim 100 -130$ mas \citep{chi99}. The published results, corrected
for Galactic extinction are plotted in Figure 2
\citep{chi99,chi02,pri16}. The data is non-simultaneous and is
distributed from 1995 - 2003. The variability implied by the scatter
indicates a factor of $<2$ variability over time. 130 mas is large
compared to the 0.040 mas EHTC. However, inspection of 5 GHz,
\citet{had14}, 15 GHz, \citet{lis13}, 43 GHz, (\citet{mer16,had14}
and 86 GHz, \citet{had16,kim16}, VLBI images indicate that there is
no strong optically thick component between 0.06 mas and 100 mas
that could produce a significant contribution to the opticl/UV flux
density - the unresolved 86 GHz core is the only possible source of
the HST detected flux. We can get a tight bound on the high
frequency synchrotron tail by assuming that most of this emission is
associated with the smaller EHTC. This assumption has been made
previously \citep{dex12}. These data are plotted in Figure 2 as well
as the 11.7$\mu$ Mid-IR flux density with 400 mas resolution which
provides a loose upper bound on the synchrotron tail in the gap in
spectral coverage \citep{why04}.

\subsection{Explicit Models for the Spectral Fits to the EHT Core}
The homogeneity produced by the volumetric averages described at the
end of Section 2 simplifies the radiative transfer equation,
allowing for a solution of the form of Equation (4) for every path
through the plasmoid with the same $\mu(\nu)$, $j_{\nu}$ and
$\delta$. The various models are described in Table 1. Each ``model"
has a preassigned $\alpha$, $a$, $v^{z'}$ and $H/R$ (the last four
columns in Table 1) plus the $\rm{LOS}$, $\psi=15^{\circ}$. Each
model has a corresponding fit to the data in Figure 2 indicated in
columns 2 - 4 in Table 1 by the three flux densities). The term
model represents an infinite number of degenerate solutions as
indicated by the curves in Figures 3 and 4. Due to the homogeneous
approximation and the volumetric averages described at the end of
Section 2, the spectrum will depend on the uniform values of
$\mu(\nu)$, $j_{\nu}(\nu)$, $\delta$. $\psi$, $\alpha$, $H$ and $R$;
there are 7 parameters. The models have the 5 preassigned values and
three free variables $N$, $B^{P}$ and $R$.\footnote{Formally, the
variable, $N_{\Gamma}$ that is defined in Equation (3), is a
surrogate for the variable, $N$ since Equations (1) and (5) only
depend on $N_{\Gamma}$. As discussed in reference to Equation (3),
since the low energy cutoff cannot be determined there is a
significant uncertainty in $N$ and results depending on $N$ are not
well constrained and therefore not considered in this study.} Thus,
8 model values are used to solve for the 7 parameters that determine
the tubular jet spectrum. Eliminate the preassigned values of the
models that are common for the description of the jet spectrum
($\psi$, $\alpha$ and $H/R$). The problem reduces to 5 model values
($N$, $B^{P}$, $R$ are free to vary and $a$ and $v^{z'}$ are fixed)
that determine the 4 parameters required to generate the spectrum
($\mu(\nu)$, $j_{\nu}(\nu)$, $\delta$ and $R$). Thus, in each model
class, there are actually an infinite number of physical solutions
for the same fit, the one dimensional curves in Figures 3 and 4.
There are 15 models and 9 different fits. As a consequence of the
single values of $\mu(\nu)$, $j_{\nu}(\nu)$ and $\delta$ throughout
the calculational volume, the models A-E have been chosen to have
exactly the same fit to the data. This fit is a control variable in
the numerical experiments to follow. The data are upper limits.
Thus, the fits in Figure 2 explore a wide range of the excess of the
data relative to the actual flux produced by the forward jet base
located inside the EHTC.
\par In order to interpret $F_{\nu}(\nu_{0} = 230 \, \rm{GHz})$, note that simulated models indicate that a gravitationally lensed counter jet
and/or the accretion disk itself can produce the observed
$F_{\nu}(\nu_{o} = 230 \, \rm{GHz})$ \citep{dex12,mos16}. In the
models, a luminous disk will have an opacity sufficient to absorb
the lensed counter jet emission \citep{dex12}. In these models, the
pressure-driven "funnel wall jet" initiates in a distributed region
suspended along the interface between the accretion vortex and the
disk, i.e., at larger cylindrical radii than the ISCO (innermost
stable orbit) and above the equatorial plane \citep{haw06,mos16}. By
contrast, in the present model, the jet initiates just outside the
ISCO in the equatorial plane, assuming the role of the luminous
inner disk, thereby plausibly absorbing the lensed counter-jet
emission. Thus, the forward tubular jet base can be the predominant
source of $F_{\nu}(\nu_{o} = 230 \, \rm{GHz})$. However, because of
the uncertainty in the source of EHT correlated flux, the parametric
study presented here allows for the disk and counter jet to produce
a wide range of $F_{\nu}(\nu_{o} = 230 \, \rm{GHz})$, from 0 to 2/3
of the total.

\begin{figure}

\includegraphics[width= 0.85\textwidth]{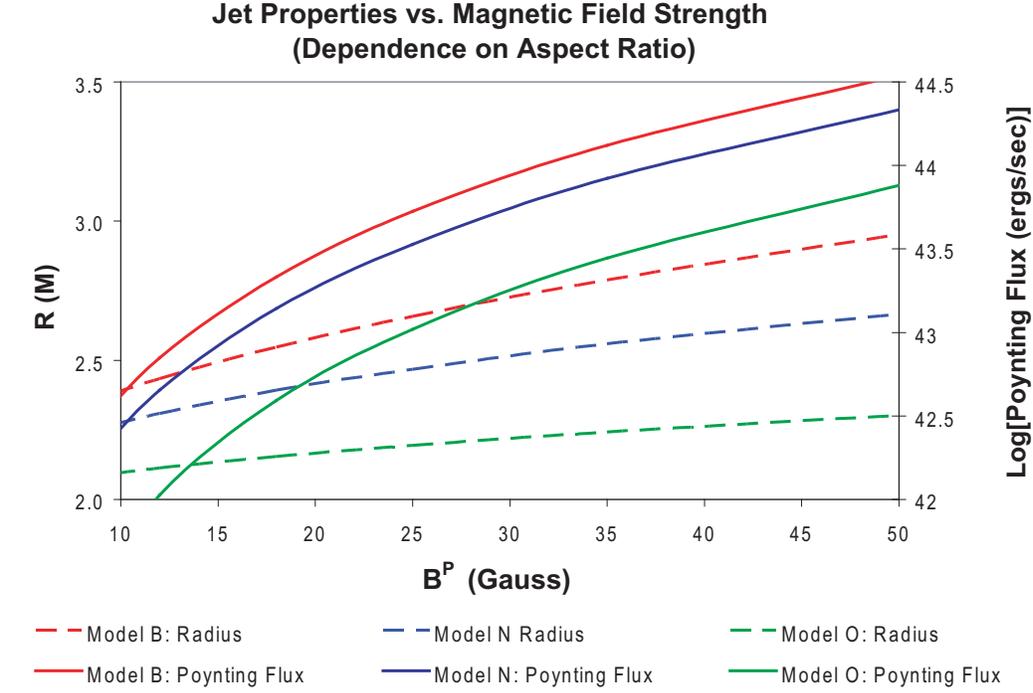}
\includegraphics[width= 0.85\textwidth]{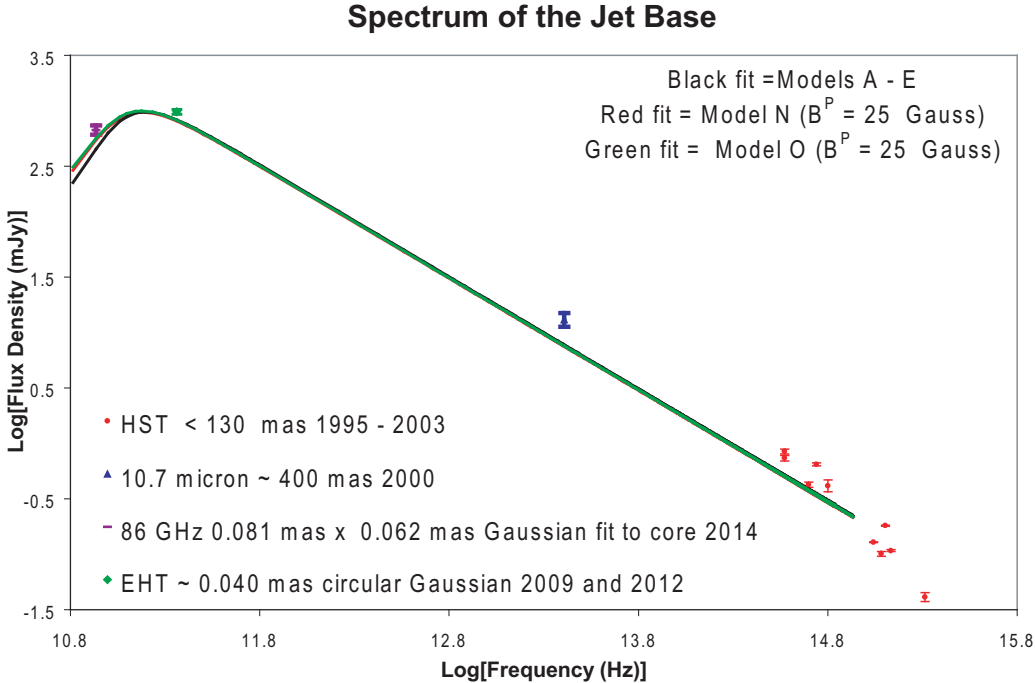}

\caption{The dependence of outer radius, R (top), Poynting flux
(top) and spectrum (bottom) on the aspect ratio $H/R$, where $H$ is
the jet length. Models B, N and O have $H/R=2$, $H/R=4$ and $H/R=8$,
respectively. All models have $a/M=0.95$, $v^{z'}=0.1$,
$\alpha=1.01$ and the same optical flux. The larger the aspect
ratio, the narrower the tubular jet. $R$ must decrease in order to
maintain a similar volume, otherwise the jet will over-produce
optical emission relative to the fiducial Model B (bottom frame).
The smaller $R$ also reduces the crosse-sectional area of the jet
and therefore, the Poynting flux at fixed $B^{P}$. }

\end{figure}

\par Table 1 lists the 15 models that are used to analyze the tubular
jet model in this study. There are four separate issues that are
being investigated by the parametric study
\begin{enumerate}
\item The black hole spin parameter, $a/M$
\item The axial velocity of the jet, $v^{z'}$
\item The spectral fit to the uncertain broadband spectrum of the
jet base discussed in the last subsection
\item The length, $H$, of the jet base responsible for the broadband
spectrum given by the aspect ratio $H/R$.
\end{enumerate}
In order to explore each of these items, three items should be held
fixed, with the fourth allowed to vary.
\par Most of the models have $H/R=2$. A larger $H/R$ might be more physically reasonable for a jet base, but a simple uniform
right circular tube plasmoid model is less justified. In the models,
the tubular plasmoids have a length, $H\approx 5M \approx 4.2\times
10^{15} \,\rm{cm}$. For an average jet propagation speed of 0.1c,
this corresponds to a propagation time for an element of plasma to
traverse a jet base length, $H$, of $t_{jet}=1.4 \times
10^{6}\,\rm{sec}$. Based on the gravitational redshift, $\nu_{o}=
230\, \rm{GHz}$ corresponds to $\nu \approx 300 \, \rm{GHz}$. The
synchrotron lifetime at the peak emission frequency, $\nu_{m}$, in
the plasma rest frame is \citep{tuc75}
\begin{equation}
t_{\rm{sy}} \approx \frac{5\times 10^{11}}{(B^{3}\nu_{m})^{1/2}}\;
\rm{sec}\;.
\end{equation}
Using $B=25 \rm{G}$,  $t_{jet}/t_{\rm{sy}} \approx 190$. Thus
$H/R=2$ seems short, but dynamically it is a long time. The basic
premise of this analysis is that when the jet is ejected from the
hot denser accretion flow it is highly luminous and it is this jet
base that is being modeled. The parametric analysis below will
consider the possibility that much of the 86 GHz and 230 GHz
emission might arise farther out in the jet. Addressing the
uncertainty in the flux density of the jet base is the basic
principle of this parametric study. It does not make sense to model
a large region of the jet by a single zone model and that is not the
intent here.

\subsubsection{Exploring Spin Variation, Fiducial Fit and Fiducial Model}
Models, A-C, explore changes in the jet as the spin is varied. Per
point 2), above, the axial velocity is fixed at $v^{z'}=0.1c$. The
nonrelativistic value is motivated by high resolution VLBI component
motion discussed in the last section \citep{mer16,had17}. Per point
3), above, a fiducial fit is chosen for comparison purposes. It is
arbitrary, since the data do not constrain the choice that strongly.
The fit assumes that most of the EHTC $F_{\nu}(\nu_{o} = 230 \,
\rm{GHz})$ is attributed to the base of the forward jet,
$F_{\nu}(\nu_{o} = 230 \, \rm{GHz})= 830\;\rm{mJy}$. Secondly, due
to the core shift analysis discussed in the previous subsection, the
230 GHz core is only $\sim10\mu\rm{as}$ from the center of the
Gaussian fit to the 86 GHz core. Combining this with
$F_{\nu}(\nu_{o} = 230 \, \rm{GHz})= 830\;\rm{mJy}$, and assuming
that the spectrum has only weak SSA absorption at 230 GHz, a
substantial flux density at 86 GHz is expected. It must be less than
the total flux fit in \citet{had16}, thus a value of
$F_{\nu}(\nu_{o} = 86 \, \rm{GHz})= 450\,\rm{mJy}$ is chosen. The
other constraint on the fit comes from the HST observations, where
we used VLBA observations to argue that it is a tight upper limit to
the EHTC flux: $F_{\nu}(\nu_{o} = 3.72 \times 10^{14} \, \rm{Hz})=
0.5\;\rm{mJy}$.

\par Per point 4, above, a fiducial value of $H/R=2$ is chosen
based on the simplifying assumption of a uniform small region, yet
it still provides an elongated aspect of a jet. By Equation (19),
this a dynamically significant length of jet to consider.

Note that by construction, Models A-C have the same spectrum (see
Figures 2a and 3a). For most of the comparative analysis to follow,
Model B is utilized as a fiducial model of the physical state of the
system, $v^{z'}=0.1c$, $a/M=0.95$, and $H/R=2$. Figure 3a shows a
larger radius for lower spin. This is expected since the ISCO is
farther out. The Poynting flux is larger at a fixed $B^{P}$ for
higher spin. This is primarily because Equation (18) indicates a
quadratic dependence of the Poynting flux on $\Omega_{F} \approx
\Omega_{\rm{kep}}(r_{o})$ and by Equation (9), the smaller ISCO for
higher spin indicates a much larger $\Omega_{\rm{kep}}(r_{o})$
throughout the base of the tubular jet.
\subsubsection{Exploring Axial Velocity Variation}
The next set of models, D and E, explore the effects of varying the
axial velocity, $v^{z'}$. In this case not only must the fit and
$H/R$ be fixed, but also the spin. In this regard, $a/M=095$ is
chosen as the fiducial physical state of the black hole and Model B
is the fiducial model for comparison. Note that models A-E will have
the same spectrum by construction since this is a control variable
in the numerical experiment (see Figures 2a and 3b).
\par {The axial velocity is necessarily nonrelativistic if the
outflow has bilaterl symmetry: this implies $v^{z'}=0$ at the
equator. This is consistent with the VLBI observations of component
motion within 2 mas of the BH that indicates speeds on the order of
0.1c - 0.4c \citep{had16,had17,mer16}. The slowest jet has the
smallest radius, yet the overall variation between $v^{z'}=0.05$ and
$v^{z'}=0.2c$ is only 2\% - 3\%. The Poynting flux is smaller for
$v^{z'}=0.05$ at a fixed $B$ and $a$ due to the smaller radius and
therefore less cross-sectional surface area in the integral of
Equation (18). The variation between $v^{z'}=0.05$ and $v^{z'}=0.2c$
is minimal.
\subsubsection{Exploring Variation in the High Frequency Synchrotron Tail}
\par Models F and G hold properties 1,2 and 4, above, constant. The experiment
considers variations in the fit if $F_{\nu}(\nu_{o} = 3.72 \times
10^{14} \, \rm{Hz})$ is allowed to vary, with $F_{\nu}(\nu_{o} = 86
\, \rm{GHz}) = 450\;\rm{mJy}$ and $F_{\nu}(\nu_{o} = 230 \,
\rm{GHz})= 830\;\rm{mJy}$ held constant. When combined with the
fiducial Model B, this is essentially an exploration of the
dependence of the physical parameters of the tubular jet model
solely based on the variation in the synchrotron tail (see Figures
2d and 4d).
\par There is very little change in $R$ and the Poynting flux even
with a factor of 2 change in the optical flux. Thus, exact knowledge
of the optical flux from the EHTC is not necessary for an accurate
estimate of the size and jet power in the tubular jet model.
\subsubsection{Exploring the Uncertainty in the 86 GHz Flux Density}
\par Models H and I hold properties 1,2 and 4 above constant. This numerical experiment
considers variations in the fit to the data if $F_{\nu}(\nu_{o} = 86
\, \rm{GHz})$ is allowed to vary, with $F_{\nu}(\nu_{o} = 3.72
\times 10^{14} \, \rm{Hz})= 0.5\;\rm{mJy}$ and $F_{\nu}(\nu_{o} =
230 \, \rm{GHz})= 830\;\rm{mJy}$ held constant. This is essentially
an exploration of SSA opacity variation when combined with the
fiducial Model B. It is motivated by the fact that the amount of
$F_{\nu}(\nu_{o} = 86 \, \rm{GHz})$ that is attributable to the jet
base is uncertain as discussed in Section 3.1 (see Figures 2a and
4a).
\par {A large variation in the tubular jet model is seen based on the
assumed value of $F_{\nu}(\nu_{o} = 86 \, \rm{GHz})$ from the jet
base. Lower $F_{\nu}(\nu_{o} = 86 \, \rm{GHz})$ with
$F_{\nu}(\nu_{o} = 230 \, \rm{GHz})= 830\;\rm{mJy}$ held fixed means
a higher opacity. At a fixed $B^{P}$, by Equation (1), this can be
achieved with a higher $N$ or $R$: $\mu(\nu)\sim NH\sim NR$.
However, the luminosity of the synchrotron tail $F_{\nu}(\nu_{o} =
3.72 \times 10^{14} \, \rm{Hz})\sim NHR^{2}\sim NR^{3}$. This is
held fixed, so the only solution is an increase in $N$ and a
decrease in $R$. The higher opacity solutions have larger $N$ and
smaller $R$. For a fixed $B^{P}$, $v^{z'}$ and $a$, Equation (18)
indicates a smaller Poynting flux as well due to the smaller
cross-sectional area of the jet associated with the smaller $R$
values. The conclusion is that the uncertainty in $F_{\nu}(\nu_{o} =
86 \, \rm{GHz})$ presents significant uncertainty in the tubular jet
models.
\subsubsection{Exploring the Uncertainty in the 230 GHz Flux Density}
\par Models J and K hold properties 1,2 and 4 above constant. This numerical experiment
considers variations in the spectral fit if $F_{\nu}(\nu_{o} = 230
\, \rm{GHz})$ is allowed to vary with $F_{\nu}(\nu_{o} = 3.72 \times
10^{14} \, \rm{Hz})= 0.5\;\rm{mJy}$ and $F_{\nu}(\nu_{o} = 86 \,
\rm{GHz})= 450\;\rm{mJy}$ held fixed. This is essentially an
exploration of peak spectral flux density variation when combined
with Model B. It is motivated by the fact that $F_{\nu}(\nu_{o} =
230 \, \rm{GHz})$ from the jet base is uncertain based on the
discussion at the beginning of Section 3.2 (see Figures 2b and 4b).
\par Surprisingly, changing $F_{\nu}(\nu_{o} = 230
\, \rm{GHz})$  has little effect on the tubular jet models.
Ostensibly, this appears to be another opacity study, but with
different results than Section 3,2.4. There is a significant
difference from Section 3,2.4. The synchrotron luminosity near the
peak and throughout the sub mm is not held fixed. Every increase in
opacity corresponds to an increase in the synchrotron luminosity.
The opacity and total luminosity affect $R$ in an opposite sense.
The effects on the tubular jet model are opposite and tend to cancel
out.

\begin{figure}
\includegraphics[width= 0.85\textwidth]{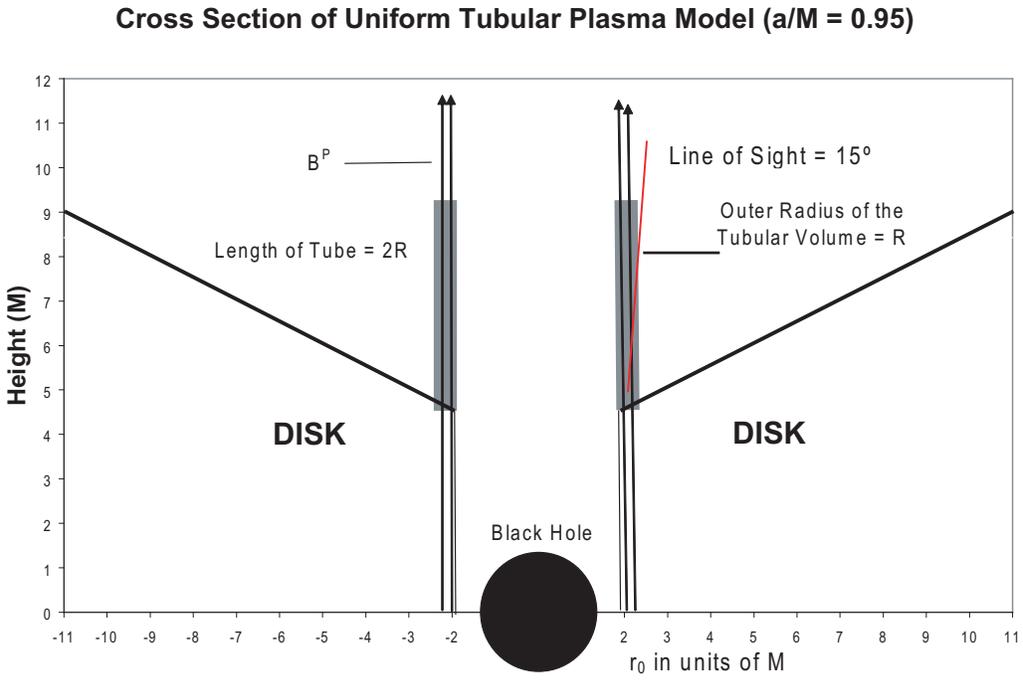}

\caption{The jet need not begin at the equatorial plane. For
example, it can begin above the accretion disk. The field lines are
still anchored in the equatorial plane. The example chosen here has
$H/R=2$ and $a/M=0.95$ as in Figure 1.}
\end{figure}
\subsubsection{Exploring a Small Jet Contribution to the 230 GHz Flux Density}
\par Models L and M hold properties 1,2 and 4 above constant. This study
considers the possibility that the forward jet base flux density is
a small fraction of the total observed EHTC flux density. For
example, the majority of the flux density might be from the jet much
farther out than the model of the jet base, the disk or the
counter-jet. Combining these models in combination with Model I give
a second parametric study of the variations in the fit if
$F_{\nu}(\nu_{o} = 230 \, \rm{GHz})$ is allowed to vary with
$F_{\nu}(\nu_{o} = 3.72 \times 10^{14} \, \rm{Hz})= 0.5\rm{mJy}$ and
$F_{\nu}(\nu_{o} = 86 \, \rm{GHz})= 350\rm{mJy}$ held constant (see
Figures 2c and 4c). This is a low flux version of the peak flux
variability study involving Models B, J and K in Section 3.2.5. The
results of the numerical experiment is depicted in Figures 2c and
4c.
\par This numerical experiment replicates the result of the preceding
case. Very little change is seen in the tubular jet model as the 230
GHz flux density is varied. It shows that tubular jet model is not
strongly perturbed even if the majority of the EHT correlated flux
density is not attributable to the forward jet base. The surprising
result of these two experiments indicate that exact knowledge of the
230 GHz flux density from the EHTC is not required to constrain the
tubular jet model.
\subsubsection{Exploring Variations in the Length of the Jet Base}
\par The final parametric study holds properties 1) - 3), above,
fixed but vary the aspect ratio of the jet, $H/R$. Models N and O
are identical to Model B in terms of properties 1) - 3). However,
$H/R$ is changed to 4 and 8 in Models N and O, respectively, as
opposed to 2 in Model B (the results are in Figures 5a and 5b).
\par The fiducial jet length of $2R$ is motivated by considering
the plausibility of a uniform tubular geometry. As the tube gets
longer, the right cylindrical shape and uniformity become less
accurate descriptions and the model becomes more complicated.
Besides the shape and uniformity, major concerns are posed by the
effects caused by the change in the redshift of Equation (17) as the
jet gets farther from the black hole. The short $H/R=2$ tube is a
crude one zone model used as an approximation in order to explore
basic parameter changes. Extrapolating a single zone model that is
designed to explore the base of the jet to a many-fold longer jet
length is not justified. Thus, a longer jet is built up by
connecting shorter $H/R=2$ length modules end to end. Each module
has its own volumetric averages of $\delta^{2.5 +\alpha}$ and
$\gamma_{\rm{Kep_Z}} (\delta^{3 +\alpha})$ per the strategy
described at the end of Section 2. The plasma properties, $N$,
$B^{P}$, $\alpha$ and $R$ are identical from module to module.

\par The top frame of Figure 5 shows the dependence of $R$, and Poynting flux
on the aspect ratio $H/R$. All models have $a/M=0.95$, $v^{z'}=0.1$,
$\alpha=1.01$ and the same optical flux. The larger the aspect
ratio, the narrower the tubular jet becomes. $R$ must decrease in
order to maintain a similar volume, otherwise the jet will
over-produce optical emission relative to the fiducial Model B
(bottom frame of Figure 5). There is maximal redshift in the bottom
tubular segment and therefore less SSA absorption and less
synchrotron luminosity. The upper tubular segment is the most
dominant contributor to the luminosity and has the highest SSA
opacity (see Equations (1) and (5)). At $H/R=8$, the extrapolation
has exhausted the effects of gravitational redshift. There is a
small difference between the third segment and the fourth segment.
For example, $F_{\nu}(\nu_{o} = 3.72 \times 10^{14} \, \rm{Hz})$ of
the segment from $4R$ to $6R$ is $\gtrsim 85\%$ of $F_{\nu}(\nu_{o}
= 3.72 \times 10^{14} \, \rm{Hz})$ of the segment from $6R$ to $8R$.
Likewise, the SSA optical depth at 140 GHz is 0.89 in the third
segment and 1.04 in the fourth segment. Creating additional length
per this method becomes stacking more similar segments with the same
opacity and the same luminosity per unit length. The net result is
to shrink the radius to maintain a similar volume. The long,
thin-walled, tube when $H/R=8$ seems unrealistic. One is probably
not exploring physical changes in the long tubes, but seeing a break
down of the assumptions of the model.
\par The bottom frame of Figure 5 shows the resultant spectra from the segmented models.
Since the opacity is different in every segment, the fiducial fit of
Model B cannot be attained. The models were an attempt to get close
to the same spectrum as Model B, so that a comparison can be made.
The lower opacity of the base results in an excess of flux density
at 86 GHz for models N and O relative to model B if the optical flux
and the 230 GHz flux density are held approximately equal to that of
Model B. Note that an exact value of $B^{P}$ is chosen for the
spectra of Models N and O. Each value of $B^{P}$ produces a slightly
different spectra, so a particular representative value was plotted.
\par In spite of not being able to produce identical spectral
fits, the models demonstrate one basic conclusion: the largest
angular size transverse to the jet direction occurs for the shortest
models (in this case $H/R=2$). The philosophy of this model was to
assume that the base of the jet emerging from the hot accretion flow
is very luminous. Depicting this emission near the BH is the intent
of the basic model, not emission farther out along the jet. Future
EHT observations may be able to resolve flux on the scales
indicative of the jet base.

\subsubsection{Exploring Variations in the Elevation of the Jet Initiation Point}
{Another variable in the basic configuration is the elevation of the
launch point above the equatorial plane. The outgoing jet might be
lower density material that initiates at the top boundary of a
denser accretion disk as depicted in Figure 6. The $H/R=2$ jet that
initiates 2R above the equator has the same properties 1) - 4) as
Model B. The lone difference is the elevation of the initiation
point. This results in a less than 1\% decrease in $R$ and the
Poynting flux. This negligible change arises from a near cancelation
of effects. Note that by Equation (1), $\mu(\nu_{o}) \sim
NH\delta^{2.5 +\alpha}$, and from equations (5), (7) and (12), the
synchrotron luminosity, $L(\nu_{o}) \sim NHR^{2} \gamma_{\rm{Kep_Z}}
(\delta^{3 +\alpha})$. The change in elevation results in a change
to the gravitational redshift contribution to $\delta$ in Equation
(17) since the volume is farther from the BH with less redshift.
This effect is surprisingly small since $\gamma_{\rm{Kep_Z}}
(\delta^{3 +\alpha})/(\delta^{2.5 +\alpha}) \approx
\gamma_{\rm{Kep_Z}} (\delta^{0.5})\approx 1$ throughout the volume
of both the elevated jet and the equatorial plane launched jet and
likewise so are the volumetric averages that are used.
\par Properties related to the number density are not plotted because
there is a huge uncertainty since the low energy cutoff in Equation
(3) is unconstrained by the insufficient resolution of VLBI at cm
wavelengths. However, the change in elevation of the initiation
point and the resultant smaller gravitational redshift actually
decreases the proper number density, $N$, by a factor of $\sim 4$.
Elevated jet initiation might be physically more reasonable and more
conducive to establishing jet solutions that are magnetically
dominated. Recall the discussion that motivates Equation (14)
indicating that the jet begins highly magnetically dominated in
order to explain the observed acceleration on sub mas scales.

\section{Conclusion} EHT observations at 230 GHz were combined with
86 GHz VLBI observations in order to constrain the SSA opacity.
Considering 0.1" resolution HST optical photometry in the context of
VLBI images of the jet on scales $\leq 100\,\rm{mas}$ indicates that
the EHTC is the most plausible source of the HST flux. These data
indicate a large SSA opacity at $\sim 100$ GHz and a modest
IR/optical synchrotron (HST) luminosity. These constraints are
applied to the tubular jet base models illustrated in Figure 1.

\par Section 3 is a parametric analysis of possible tubular jet
models that are consistent with the data. 15 models are considered
corresponding to 9 different fits to the data (see Table 1 and
Figure 2). Due to insufficient resolution and/or lack of imaging,
the observations do not tightly constrain the the flux density of
the EHTC. To compensate for the uncertainty, many fits to the data
and models were explored. For each model, $R$ and the Poynting flux
are plotted as a function of the proper poloidal magnetic field,
$B^{P}$, in Figures 3 and 4. In order to interpret the results
displayed in Figures 3 - 4, it is useful to have an expectation on
the Poynting flux. Isotropic estimates of jet power yield $Q = 0.75
- 6 \times 10^{43} \rm{ergs/s}$ \citep{mcn11,wil99,pun05}. Estimates
based on the brightest features in the interior jet are biased
towards the more energetic episodes in the jet history and find
$Q\gtrsim 10^{44} \rm{ergs/s}$ \citep{sta06,owe00}. A reasonable
range is $Q\sim 10^{43}-10^{44} \rm{ergs/s}$. The parametric
analysis of the jet base indicates the following.

\begin{enumerate}
\item The entire power budget of the M87 jet can be accommodated for
$0.99> a/M > 0.95$ ($a/M\sim 0.7 $) if the outer radius of the
emitting region is $\sim 1.8 - 3.0 M$ ($\sim 3.5 - 3.9 M$) and the
vertical magnetic field is 8 - 40 G ($20- 50$ G).
\item The dimension transverse to the jet direction is 12-21$\,\mu\rm{as}$ ($\sim 24 - 27\, \mu\rm{as}$) for $0.99 > a/M> 0.95$ ($a/M\sim 0.7 $), $\approx 1/2$
the EHTC size in the numerical models of \citet{dex12,mos16}. If the
notion of a compact luminous jet base near the black hole is
abandoned and aspect ratios, $H/R>2$, are considered, the minimum
transverse sizes are decreased. EHT imaging might be able to
discriminate between the models.
\item The jet base dimensions and power depend strongly on $F_{\nu}(\nu_{o} = 86 \, \rm{GHz})$
and weakly on other parameters such as jet speed, $F_{\nu}(\nu_{o} =
230\, \rm{GHz})$ and optical flux.
\end{enumerate}
This analysis is not a full radiative transfer calculation that
captures effects such as gravitational lensing \citep{dex12,mos16}.
It relies on the simplified solution of Equation (4). The analysis
does incorporate gravitational redshift and transverse Doppler
shift. The preferred configuration was argued in Section 3.2.8 to be
a jet that initiates at the top of the disk as illustrated in Figure
6. It would be interesting to see if there are some significant
changes to the apparent transverse size in this preferred
configuration if the potential effects of gravitational lensing are
included.
\par The leptons responsible for the EHTC emission will
synchrotron and possibly self-Compton cool (see the discussion of
Equation (19)) near the BH and must be reheated in order to explain
the almost hollow jet detected in \citet{had16} on 0.1 - 0.5 mas
scales at 86 GHz. The tubular jet model provides a natural
explanation of this emission at a de-projected distance of $\sim
120M - 600M$ from the central BH. The jet is Poynting flux
dominated. It carries a large, $\sim 10^{44} \rm{ergs/sec}$, energy
flux outward in a tubular conduit, the tubular jet. It only takes a
negligible fraction of the Poynting flux to be converted to leptonic
heating in order to repeatedly re-energize the plasma, thereby
making it synchrotron luminous. There are three common mechanisms
for heating the plasma. The first mechanism are shocks created near
the tubular jet boundary. Even though these fast shocks are not that
efficient for heating the plasma as noted in \citet{ken84}, they
need not be because there is a large Poynting flux reservoir
available for this process. Particle-in-cell simulations have
indicated that high energy particles can be created by a variety of
processes in shocks, including surfing acceleration on strong
electrostatic waves in nearly perpendicular shocks \citep{mat17}.
Another likely possibility is reconnection. Field tangling is often
called braiding in solar physics. Braided fields are believed to
release the extra energy of tangling as they relax to a more
simplified state by reconnection \citep{wil10}. Reconnection of the
braided fields in the jet can also provide high energy plasma to the
jet and the fields are strongest (the most stored energy) in the
tubular shell of the jet \citep{wil10,bla15}. Thirdly, there is
almost certainly a turbulent shear layer at the outer boundary of
the jet as it transitions to the intergalactic medium. The
reconnection of magnetic turbulence is also a source of high energy
particles \citep{laz15}. All three of these process are likely at
work, repeatedly injecting new plasma at irregular intervals. This
naturally explains the fact that the brightest features seem to be
in a different place at different epochs of 86 GHz VLBI observations
\citep{kim16,had16}. Reheating is consistent with particle
acceleration processes that need not be stationary in time and space
as opposed to a smooth continuously cooling hot jet.

\par The tubular models benefit from new EHT and/or 86 GHz VLBI data
that was not available in earlier models
\citep{abd09,bro09,kin15,ghi05,tav08}. The large MHD Poynting flux,
$S^{P}$, in the tubular jet models is contrary to the spine-sheath
jet models of \citet{ghi05,tav08} in which most of the jet power is
in the spine. They resemble the original spine-sheath models of jets
from rapidly rotating BHs in which the preponderance of $S^{P}$
resides in the outer jet \citep{pun96,pun08}. {An interesting
variant of the model of a strong tubular jet and a weak event
horizon driven spine is the scenario in which the preponderance of
the jet emission is emitted from just inside the ISCO, an
``ergospheric disk jet." Three dimensional MHD simulations of
ergospheric disk jets have been discussed in detail for $a/M=0.99$
\citep{pun09}. These jets can be stronger sources of Poynting flux
than jets that form outside of the ISCO.

\par This study shows that a tubular jet from the inner accretion
flow can be the source of energy that powers both the EHT core and
the large scale jet, including the energy flux needed to power
extreme dissipation sites such as HST-1 \citep{sta06}. As such,
there is no energetic requirement for a powerful spine, however it
does not disprove the possibility of a powerful spine. The
observational advantage over considering a powerful spine is that
models that drive the spine from the event horizon produce an
invisible forward jet due to the lack of energetic plasma at the jet
base. The strong jet spine models initiate within the accretion
vortex that is almost devoid of plasma. Thus, the claim that there
is insufficient energy contained within the particles in the jet
base that can be released as radiation \citep{mos16}. Thus, the
powerful jet spine is never a significant contributor to the
correlated EHTC flux density at 230 GHz in current models. This is
in contradistinction to the tubular jet models that emanates from
the inner disk which is rich in hot plasma.

\par It is also noted that the spine might be relatively weak as in
the the older spine-sheath type models \citet{pun96,pun08}. There is
currently no direct observational evidence indicating that most of
the jet energy is in the spine. Observational evidence of a spine
has been seen with VLBI at 5-15 GHz \citep{asa16,had18}. A clear
central ridge was resolved with 15 GHz VLBI at a distance of $\sim
13.5- 30$ mas, a de-projected distance of $>17000 M$ from the BH
\citep{had18}. The central ridge is less luminous than the outer
sheath, especially the southern ridge. It cannot be traced back to
the source and seems to merge with the southern ridge $\sim 13.5$
mas from the BH. Furthermore, the central ridge does not seem to
"light up" until right after (based on distance from the BH) the
outer ridges brighten, suggesting that the feature might be
generated by the outer ridges themselves. The total fraction of the
15 GHz luminosity of the jet within 30 mas that is produce by the
central ridge is negligibly small. In summary, there is no
observational evidence indicating a powerful spine of plasma being
emitted from the black hole. There is an explanation of the faint
ridge as arising from the tubular jet. The central axis is a natural
place for shocks from the outer boundary to coalesce creating an
axial region of enhanced dissipation \citep{san83}.
\par The tubular jet model offers
observers a more tangible set of predictions than an invisible
powerful spine jet. This is a strong motivation for pursuing models
in which the outer sheath is the major source of energy flux from
the black hole accretion system. The flux detected by observers can
then be used to directly constrain the energetics and microphysics
of the jet form the EHTC to large distances.
\begin{acknowledgements}
I am grateful to Donato Bini for helping me with early versions of
this manuscript. I would also like to thank an anonymous referee for
many valuable and insightful comments.
\end{acknowledgements}

\end{document}